\documentclass[aps,twocolumn,prl,showpacs,preprintnumbers,amssymb,english]{revtex4-1}
\usepackage{mathrsfs}
\usepackage{hyperref}

\usepackage{microtype}
\usepackage{graphicx}
\usepackage{amsmath, amssymb}
\usepackage{babel}
\usepackage{color}
\usepackage{slashed}
\usepackage{lmodern}
\usepackage[T1]{fontenc}
\usepackage{tikz}
\usepackage{tikz-feynman}
\usepackage{subcaption}
\usepackage{float}

\renewcommand{\thetable}{\Roman{table}}

\def\beqn{\begin{eqnarray}}
\def\eeqn{\end{eqnarray}}
\def\barr{\begin{array}}
\def\earr{\end{array}}
\def\btab{\begin{tabular}}
\def\etab{\end{tabular}}
\def\bite{\begin{itemize}}
\def\eite{\end{itemize}}
\def\bcen{\begin{center}}
\def\ecen{\end{center}}

\newcommand{\magnitude}[1]{ \mathbf{ {#1} } }
\newcommand{\magnitudeexp}[2]{ \mathbf{ {#1} }^{ \!\!\! {#2} } }

\begin{document}

\title{An \textit{ab initio} strategy for taming the nuclear-structure dependence of $ V_{ud} $ extractions: the $ {}^{10}\mathrm{C} \rightarrow {}^{10}\mathrm{B} $ superallowed transition}

%%%%
% authors and affiliations
%%%%

\author{Michael Gennari$^{1,2}$}
\author{Mehdi Drissi$^{1}$}
\author{Mikhail Gorchtein$^{3,4}$}
\author{Petr Navr\'atil$^{1,2}$}
\author{Chien-Yeah Seng$^{5,6}$}

\affiliation{$^{1}$TRIUMF, 4004 Wesbrook Mall, Vancouver, BC V6T 2A3, Canada}
\affiliation{$^{2}$University of Victoria, 3800 Finnerty Road, Victoria, British Columbia V8P 5C2, Canada}
\affiliation{$^{3}$Institut f\"ur Kernphysik, Johannes Gutenberg-Universit\"at Mainz, 55128 Mainz, Germany}
\affiliation{$^{4}$PRISMA$^+$ Cluster of Excellence, Johannes Gutenberg-Universit\"at Mainz, 55128 Mainz, Germany}
\affiliation{$^{5}$Facility for Rare Isotope Beams, Michigan State University, East Lansing, MI 48824, USA}
\affiliation{$^{6}$Department of Physics, University of Washington, Seattle, WA 98195-1560, USA}
\date{\today}

%%%%
% abstract
%%%%

\begin{abstract}

We report the first \textit{ab initio} calculation of the nuclear-structure-dependent radiative correction $ \delta_{ \mathrm{NS} } $ to the $ {}^{10}\mathrm{C} \rightarrow {}^{10}\mathrm{B} $ superallowed transition, computed with the no-core shell model and chiral effective field theory. We obtain $\delta_{ \mathrm{NS} } = - 0.422 (29)_{ \mathrm{nuc} } (12)_{ n,\mathrm{el} } $ with a $1.6$-times reduction in the total uncertainty when compared to the current literature estimate based on the shell model and Fermi gas picture. This work paves the way for a precise determination of $V_{ud}$ from superallowed beta decays within a systematically improvable framework.

\end{abstract}

\maketitle

%%%%
% introduction
%%%%

Unitarity of the Cabibbo-Kobayashi-Maskawa (CKM) quark-mixing matrix~\cite{Cabibbo:1963yz,Kobayashi:1973fv} implies for its top-row elements, $\Delta_u\equiv|V_{ud}|^2+|V_{us}|^2+|V_{ub}|^2-1=0$. Given the experimental accuracy of these elements, the top-row CKM unitarity constraint provides an important avenue for testing the Standard Model (SM) and its extensions at low energies. A revived interest of this topic follows the recent observation of an apparent unitarity violation at $\sim 3\sigma$: $ \Delta_u = -1.48(53) \times 10^{-3} $~\cite{Cirigliano:2022yyo}. Given its high impact on the search for new physics~\cite{Belfatto:2019swo, Grossman:2019bzp, Crivellin:2020lzu, Kirk:2020wdk, Crivellin:2020ebi, Alok:2021ydy, Crivellin:2021bkd,Branco:2021vhs, 
Crivellin:2022rhw, Belfatto:2021jhf, Belfatto:2023tbv, Cirigliano:2021yto, Cirigliano:2023nol, Dawid:2024wmp}, all ingredients underlying this anomaly have to be carefully scrutinized.

From $^{10}$C to $^{74}$Rb, the 23 measured superallowed beta transitions of $T(J^P)=1(0^+)$ nuclei combine to provide currently the most accurate extraction of $V_{ud}$, the largest top-row CKM matrix element.
This extraction relies on inputs from theory as much as on experimental measurements, as summarized by the following master formula~\cite{Hardy:2020qwl}
\begin{equation}\label{sec:intro:eq:master}
    \vert V_{ ud } \vert^{ 2 }_{ 0^+ } = \frac{ \mathcal{K} }{ ft ( 1 + \delta'_{ \mathrm{R} } )( 1 + \Delta_{ R }^{ V } ) ( 1 + \delta_{ \mathrm{NS} } - \delta_{ \mathrm{C} } ) } \ ,
\end{equation}
where $ \mathcal{K} = \pi^3 \ln 2 / G_{ F }^{ 2 } m_{ e }^{ 5 } $ and the quantities $\delta_\text{R}'$, $\Delta_R^V$, $\delta_\text{NS}$ and $\delta_\text{C}$ represent SM corrections that must be computed to an absolute precision of $ 10^{-4} $ to further probe the current CKM unitarity tension. Among them, $\Delta_R^V$ and $\delta_\text{NS}$ entail the effects of the hadron/nuclear-structure dependent  electromagnetic radiative corrections (RC). Recent studies based on dispersion relations and lattice Quantum Chromodynamics (QCD) have largely pinned down the single-nucleon RC $ \Delta_{ \text{R} }^V $ ~\cite{Seng:2018qru, Seng:2018yzq, Seng:2020wjq, Czarnecki:2019mwq, Hayen:2020cxh, Ma:2023kfr}. This shifted the attention to $ \delta_{ \text{NS} } $, the nucleus-dependent part of the RC, which has traditionally been evaluated with the nuclear shell model~\cite{Jaus:1989dh, Barker:1991tw, Towner:1992xm, Towner:1994mw}. It has to be noted that splitting the full RC into the three separate pieces requires a separation of scales pertinent to Quantum Electrodynamics (QED), hadron and nuclear dynamics. Recent studies in the dispersion relation framework~\cite{Seng:2018qru, Gorchtein:2018fxl} indicated that the previously assumed scale separation was flawed. This finding led to an increased uncertainty in $ \delta_{ \text{NS} } $ which makes it currently the largest source of error in $ \vert V_{ud} \vert_{ 0^{ + } } $: $ \vert V_{ud} \vert_{ 0^{ + } } = 0.97361(5)_{ \text{exp} } (6)_{ \delta_{ \text{R}' } } (4)_{ \delta_{ \text{C} } } (28)_{ \delta_{ \text{NS} } } (10)_{ \Delta_{ R }^{ V } } $~\cite{Gorchtein:2023naa}. Hence, the reduction of the $\delta_\text{NS}$ uncertainty with \textit{ab initio} nuclear many-body methods is among the most urgent tasks for the precision test of the first-row CKM unitarity. In this Letter we report the first complete study of such kin, focusing on the $^{10}$C superallowed beta decay.

%%%%
% discussion on theory
%%%%

The key hadronic ingredient to the nucleus-independent and nucleus-dependent RCs, $\Delta_R^V$ and $ \delta_{ \mathrm{NS} } $, is the $ \gamma W $-box contribution with
\begingroup
\begin{align}\label{sec:theory:eq:box_subtraction}
    \Delta_{ R }^{ V } + \delta_{ \mathrm{NS} } = 2 \, \square_{ \gamma W }^{ b, \mathrm{nuc} } + \cdots \ \ ,
\end{align}
\endgroup
where $ \square_{ \gamma W }^{ b } ( E_e ) $ is given in Eq.~\eqref{sec:theory:eq:gW_box} with associated Feynman graph in Fig.~S1 of the Supplemental Material~\cite{SupMat}, the nuclear correction averaged over the electron phase space is given by $ \delta_{ \mathrm{NS} } = 2 \big\langle \big( \square_{ \gamma W }^{ b, \mathrm{nuc} } - \square_{ \gamma W }^{ b, n } \big) \big\rangle_{ E_e } $ and the ellipses stand for known terms not related to the $\gamma W$-box~\cite{Gorchtein:2023naa, Gorchtein:2023srs}. All hadronic physics entering the $ \gamma W $-box does so via the invariant amplitude $ T_3 $, itself derived from a generalized Compton scattering amplitude involving electromagnetic and axial charged weak currents. Explicitly, the $ \square_{ \gamma W }^{ b } ( E_e ) $ and Compton amplitude $ T_3 $ read
\begin{widetext}
\begingroup
\begin{align}\label{sec:theory:eq:gW_box}
    \square_{ \gamma W }^{ b } ( E_e ) &= - e^2 \int \frac{ d^4q }{ ( 2\pi )^4 } \frac{ M_{ W }^2 }{ M_{ W }^2 - q^2 } \, \frac{ 1 }{ q^2 + i \varepsilon } \, \frac{ 1 }{ ( p_e - q )^2 - m_e^2 + i \varepsilon } \bigg[ q^2 - p \cdot q \, \frac{ ( p \cdot q ) m_e^2 - ( p_e \cdot q ) ( p \cdot p_e ) }{ M^2 m_e^2 - ( p \cdot p_e)^2 } \bigg] \frac{ T_3( \nu, \magnitude{q} ) }{ ( p \cdot q ) \,  f_{ + } }
\end{align}
\begin{align}\label{sec:theory:eq:T3}
    T_3 ( \nu, \magnitude{q} ) & = 4 \pi i \frac{ \nu }{ \magnitude{q} }  \sqrt{ M_i M_f } \sum_{ J = 1 }^{ \infty } \, ( 2 J + 1 ) \Big\langle \Psi_f \Big\vert \bigg\{ - i T^{ \mathrm{mag.} }_{ J0 } \, ( z_f - H )^{ -1 } \, i T^{ 5, \mathrm{el.} }_{ J0 } + T^{ \mathrm{el.} }_{ J0 } \, ( z_f - H )^{ -1 } \, T^{ 5, \mathrm{mag.} }_{ J0 } \nonumber \\
    & \qquad \qquad + T^{ 5, \mathrm{mag.} }_{ J0 } \, ( z_i - H )^{ -1 } \, T^{ \mathrm{el.} }_{ J0 } - i T^{ 5, \mathrm{el.} }_{ J0 } \, ( z_i - H )^{ -1 } \, i T^{ \mathrm{mag.} }_{ J0 } \bigg\} \big( q \big) \Big\vert \Psi_i \Big\rangle \ ,
\end{align}
\endgroup
\end{widetext}
where expressions for the multipole operators $ T_{ JM } $ may be found in the Supplemental Material~\cite{SupMat}. These expressions depend on (i) the on-shell electron 4-momentum $ p_e $ (ii) the virtual gauge boson $4$-momentum $ q $ with $ \nu = q_0 $, $ \magnitude{q} = \vert \vec{q} \, \vert $, $ z_i = M_i - \nu + i \varepsilon $ and $ z_f = M_f + \nu + i \varepsilon $ (iii) the tree-level Fermi matrix element $ f_{ + } $ and (iv) the rest frame 4-momentum of the nucleus, taken in the forward limit such that $ p \approx p_i \approx p_f $ up to differences in the nuclear masses. To facilitate the use of the nuclear many-body theory, a non-relativistic reduction of the SM current operators to the effective one-body nucleonic operators~\cite{World.Sci.2004, Comp.Phys.Comm.179.2008, Atom.Nuc.Tables.23.1979} has been made by expressing the Compton amplitude in terms of the nuclear resolvent and performing a multipole expansion of the currents~\cite{Seng:2022cnq, Gorchtein:2023naa}. This formalism has been used extensively in nuclear theory to compute electromagnetic and weak transitions, e.g., see Refs.~\cite{IOP.49.2022, Phys.Rev.D.95.103011, Glick-Magid:2021uwb}. We then arrive at the expression for $ T_3 $ given in Eq.~\eqref{sec:theory:eq:T3} which further depends on (i) the multipole operators (ii) the nuclear Hamiltonian $ H = H_{ \mathrm{intrinsic} } + H_{ \mathrm{c.m.} } $ and (iii) the non-relativistic $ J^P = 0^+ $ nuclear states $ \vert \Psi_i \rangle $ and $ \vert \Psi_f \rangle $. Our conventions are consistent with the non-relativistic electroweak operator basis defined in Refs.~\cite{Comp.Phys.Comm.179.2008, Atom.Nuc.Tables.23.1979}.

\begin{figure}[t!]
\captionsetup{justification=raggedright}
\centering\includegraphics[width=\linewidth]{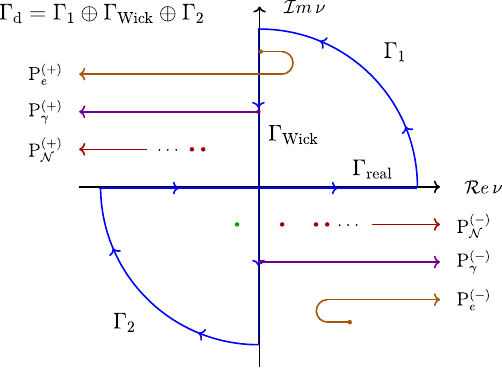}
\caption{\label{sec:theory:fig:contour_deformation} Example trajectories of the $\nu$-integral poles in Eq.~(\ref{sec:theory:eq:gW_box}) coming from the (i) nuclear propagators in $ T_3 $ (ii) photon propagator and (iii) electron propagator, with sets labelled by $ \mathcal{N} $, $ \gamma $ and $e$, respectively.}%
\end{figure}

Pragmatic evaluation of $ T_3 $ requires special attention as the resolvents, when integrated over the loop $4$-momentum in Eq.~\eqref{sec:theory:eq:gW_box}, traverse an infinite number of singularities in the discrete and continuum spectrum. Circumnavigating the vast majority is possible via Wick rotation, yet, care is required in applying Cauchy's theorem since several poles in the $\nu$-integral cross in and out of the Wick contour, as illustrated in Fig.~\ref{sec:theory:fig:contour_deformation}. Such poles may be classified into two categories: those in the electron propagator labelled by the set $ \mathrm{P}^{ (+) }_{e} $ and a subset of poles in the nuclear spectrum which involve transitions to intermediate bound states lying \textit{below} the final nuclear state, labelled by the set $ \mathrm{P}^{ (-) }_{ \mathcal{N} } $. Applying the contour deformation and accounting for incurred residue contributions, we find that
\begin{equation}\label{sec:theory:eq:gW-box_split}
    \square_{\gamma W}^{b} = \big( \square_{ \gamma W }^{b} \big)_{ \mathrm{Wick} } + \big( \square_{ \gamma W }^{b} \big)_{ \mathrm{res}, e } + \big( \square_{\gamma W}^{b} \big)_{ \mathrm{res}, T_3 } \ ,
\end{equation}
where the effect of the deformation reduces to the contribution along the $ \Gamma_{ \mathrm{Wick} } $ contour. As the Wick and electron residue terms are regular as $ E_e \rightarrow 0 $, for simplicity, we expand them to leading order in the electron energy. Overall, this leads to an $ \mathcal{O}( E_e ) $ evaluation given by
\begin{equation}\label{sec:theory:eq:gW-box_expansion}
    \square_{\gamma W}^{b} ( E_e ) = \boxminus_{ 0 } + \boxminus_{ 1 } E_e + \big( \square_{\gamma W}^{b} \big)_{ \text{res}, T_3 } ( E_e ) + \mathcal{O}( E_e^2 ) \ .
\end{equation}
Expressions for each term on the right-hand-side can be found in Ref.~\cite{Gorchtein:2023naa} as well as in the Supplemental Material~\cite{SupMat}. Analysis of the $ \square_{ \gamma W }^{b} ( E_e ) $ function requires a nuclear theory evaluation of $ T_3 $ for which we apply the \textit{ab initio} no-core shell model (NCSM)~\cite{Barrett:2013nh}.

%%%%
% contour deformation for Wick rotation
%%%%
%
%\input{contour_deformation}

%%%%
% contributions to [ gW-box-b ]
%%%%
\begin{figure*}[ht]
\captionsetup{justification=raggedright}
\includegraphics[width=\linewidth]{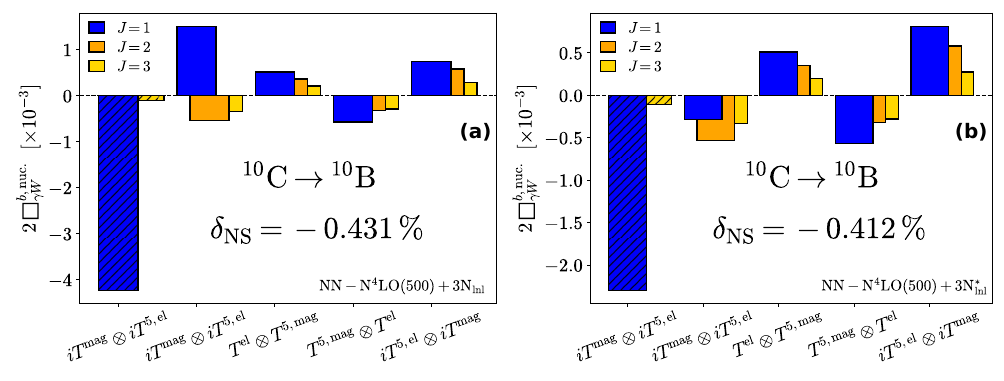}
\caption{\label{sec:results:fig:deltaNS_breakdown} Breakdown of the $ \square_{ \gamma W }^{ b,  \mathrm{nuc} } $ into (i) different electroweak operator structures in the Compton amplitude and (ii) each moment in the multipole expansion, obtained with (\textbf{a}) the chiral $\text{NN-N}^{4}\text{LO}(500) {+} 3\text{N}_{\text{lnl}}$ and (\textbf{b}) $\text{NN-N}^{4}\text{LO}(500) {+} 3\text{N}_{\text{lnl}}^{*}$ interactions, in the NCSM. Residue contributions and contributions from the electron energy expansion are shown as hatched and solid bars, respectively.}%
\end{figure*}

In the NCSM, nuclei are considered to be systems of $A$ non-relativistic point-like nucleons interacting via realistic chiral Effective Field Theory ($\chi$EFT) two-nucleon (NN) and three-nucleon (3N) interactions which serve as the sole input to the approach. Each nucleon is an active degree of freedom and the Galilean invariance of observables, as well as nuclear spin and parity, are conserved. The many-body eigenstates are expanded over a basis of antisymmetric $A$-nucleon harmonic oscillator (HO) states parameterized by the oscillator frequency $ \Omega $ and containing states with excitations of quanta up to $ N_{ \mathrm{max} } \hbar \Omega $ above the lowest Pauli configuration. The NCSM spectrum then contains both the bound discrete spectrum and a discretization of the continuum spectrum.

In this work, we apply two interactions derived in the $ \chi $EFT framework. At the two-body level, we use the $ \mathrm{NN}{-}\mathrm{N}^4\mathrm{LO}(500) $~\cite{Entem:2017gor} interaction across the board, whereas at the three-body level we apply the (a) $ 3 \mathrm{N}_{ \mathrm{lnl} } $ interaction~\cite{Gysbers2019} and (b) $ 3 \mathrm{N}^*_{ \mathrm{lnl} } $ interaction~\cite{Phys.Lett.B.845.138156}. The latter contains an additional sub-leading contact ($E_7$) enhancing spin-orbit strength~\cite{Girlanda:2011fh}. To accelerate convergence with respect to the size of the many-body configuration space, the chiral interactions are softened via Similarity Renormalization Group (SRG)~\cite{Bogner2007}. Presently, we use the evolution parameters of $ \lambda_{ \mathrm{SRG} } = 1.8 \ \mathrm{fm}^{-1} $ and $ \lambda_{ \mathrm{SRG} } = 2.0 \ \mathrm{fm}^{-1} $ for the $ \mathrm{NN}{-}\mathrm{N}^4\mathrm{LO}(500) {+} 3 \mathrm{N}_{ \mathrm{lnl} } $ and $ \mathrm{NN}{-}\mathrm{N}^4\mathrm{LO}(500) {+} 3 \mathrm{N}_{ \mathrm{lnl} }^{ * } $ interactions, respectively. To gauge convergence, model spaces up to $ N_{ \mathrm{max} } = 7 $ with oscillator frequencies in the range of $ \hbar\Omega = 16  {-} 20 $ MeV have been applied.

Employing the NCSM with $ \chi $EFT Hamiltonians in the evaluation of $ T_3 $ requires us to compute the resolvent amplitudes up to intermediate 3-momentum transfer ($ \magnitude{q} \lesssim 500 \ \mathrm{MeV} $). A na\"{i}ve calculation is computationally infeasible, yet, via subspace techniques such as the Lanczos strength method~\cite{Jour.Phys.A.7.2120, Rev.Mod.Phys.66.763, Few.Body.33.259276}, the complexity may be reduced to solving the many-body Shr{\"o}dinger equation with a source term, i.e.,\ $ \big( z - H \big) \big\vert \Psi_{ n } \big\rangle = O_{ J M } \big\vert \Psi_{ i } \big\rangle $. This approach was previously explored in NCSM evaluations of anapole and electric dipole moments~\cite{Hao:2020, Froese:2021}. In our case, $ O_{ J M } $ corresponds to any of the operators acting on the initial state in Eq.~(\ref{sec:theory:eq:T3}) and $ \big\vert \Psi_{ n } \big\rangle $ denotes the intermediate excited states; these are confined to a subspace of the Hamiltonian characterized by the symmetry group representation of the source vector $ O_{ JM } \big\vert \Psi_{ i } \big\rangle $. Remarkably, despite the limited convergence of high-lying eigensolutions, the $ \nu $-integrals over $ T_3 $ rapidly converge with respect to the Krylov subspace truncation. This is a consequence of the exact reproduction of sum-rules when employing the Lanczos strength method~\cite{haydock1972electronic, dagotto1994correlated, bessis1975perturbative}.

While we are able to numerically evaluate the resolvent amplitudes to high-precision, in using $ \chi $EFT and the NCSM we are restricted in the physics which enters the resolvent itself. The \textit{ab initio} nuclear box diagram $ \big( \square_{ \gamma W }^{ b, \mathrm{nuc} } \big)_{ \mathrm{ab} } $ only contains contributions from nucleonic intermediate states; any non-nucleonic or scattering physics above the pion-production threshold is beyond the reach of this approach. It is then sensible to combine $ \big( \square_{ \gamma W }^{ b, \mathrm{nuc} } \big)_{ \mathrm{ab} } $ with the purely nucleonic (elastic) contribution to the single-nucleon box diagram as
\begin{equation}\label{sec:theory:eq:deltaNSmod}
    \delta_{ \mathrm{NS} } = 2 \left \{ \big( \square_{ \gamma W }^{ b, \mathrm{nuc} } \big)_{ \mathrm{ab} } - \big( \square_{ \gamma W }^{ b, n } \big)_{ \mathrm{el} } + \delta \big( \square_{ \gamma W }^{ b, n } \big)_{ \mathrm{sh} } \right \} \ ,
\end{equation}
where $ ( \square_{ \gamma W }^{ b,n } )_{ \mathrm{el} } = 1.06 (6) \times 10^{-3} $~\cite{Gorchtein:2023srs}. The last term represents contributions from energies above the pion-production threshold where the production of multi-hadron intermediate states is affected by the nuclear environment, a phenomenon referred to as ``nuclear shadowing''~\cite{Armesto:2006ph, Kopeliovich:2012kw}. This contribution to $ \delta_{ \mathrm{NS} } $ is introduced here for the first time. A precise treatment of the shadowing contribution is beyond the scope of this paper, however, we infer its size from the ``inelastic'' component of the single-nucleon box corrected for nuclear shadowing effects based on experimental data, giving $ \vert \delta( \square_{ \gamma W }^{ b, n } )_{ \mathrm{sh} } \vert < 1.2 \times 10^{-4} $. Further details may be found in the Supplemental Material~\cite{SupMat} (see also Refs.~\cite{Nachtmann:1973mr, Nachtmann:1974aj, Gross:1969jf, Kataev:1994rj, Kim:1998kia, Larin:1991tj, Baikov:2010iw, Baikov:2010je, glauber1959, Gribov:1968jf} therein).

%%%%
% discussion of results
%%%%

%%%%
% momentum distribution of [ gW-box-b ] ( |q|, E_e ) @ fixed E_e
%%%%
\begin{figure}[!t]
\captionsetup{justification=raggedright}
\centering\includegraphics[width=\linewidth]{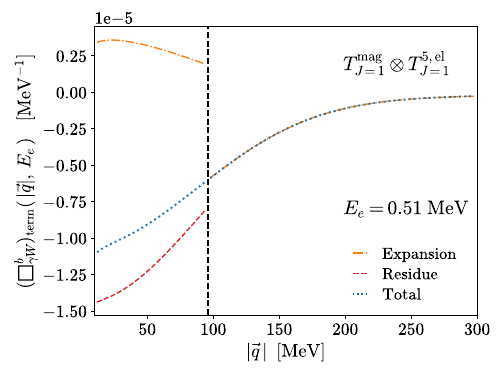}
\caption{\label{sec:results:fig:gW-box-b_momentum_distribution} Slice of $ i T^{ \mathrm{mag} }_{ J = 1 } \otimes i T^{ 5, \mathrm{el} }_{ J = 1 } $ contribution to the $ \gamma W $-box integrand at $ E_e \approx m_e $. The Compton residue (dashed), $ \mathcal{O}( E_e ) $ expansion (dash-dotted) and total sum (dotted) are as detailed in Eq.~\eqref{sec:theory:eq:gW-box_expansion}. The vertical dashed line corresponds to $ \magnitude{q}_{ \mathrm{max} } = \sqrt{ M_{ f }^2 - M_{ k }^2 } $ where $ k $ is the lowest-lying ${}^{10} \mathrm{B} $($1^+$) state. }%
\end{figure}

In Fig.~\ref{sec:results:fig:deltaNS_breakdown} we present the first \textit{ab initio} calculations of the nuclear structure radiative correction $\delta_{\text{NS}}$ obtained with the aforementioned chiral interactions. In each case, we show the breakdown of contributions to the $ \square_{ \gamma W }^{ b,  \mathrm{nuc} } $ coming from (i) different electroweak operator structures in the Compton amplitude and (ii) each moment in the multipole expansion. Interestingly, we observe a large relative size for the $ J = 1 $ Compton amplitude residue $-$ shown as a hatched bar $-$ in the total evaluation of the nuclear $ \gamma W $-box. As these residues involve transitions to the intermediate $ {}^{10} \mathrm{B} $ states lying below the final $ 0^{ + } $ excited state, i.e., they only depend on the low-lying discrete nuclear spectrum, the magnitude of the contribution can be largely attributed to the strength of $ J = 1 $ electromagnetic transitions between the $ {}^{10} \mathrm{B} $ excited $ 0^{ + } $ and lower-lying $ 1^{ + } $ states. Unlike the amplitudes involving intermediate resolvent operators, these are computationally straightforward to evaluate. However, despite the relative size of the residues being a shared feature of both calculations, we find that the distribution of strength for the $ i T^{ \mathrm{mag} }_{ J = 1 } \otimes i T^{ 5, \mathrm{el} }_{ J = 1 } $ amplitude into the residue and non-residue pieces is notably different. In the calculation with the $ \mathrm{NN}{-}\mathrm{N}^4\mathrm{LO}(500) {+} 3 \mathrm{N}_{ \mathrm{lnl} } $ interaction, the residue represents about two-thirds of the total contribution to the $ \square_{ \gamma W }^{ b,  \mathrm{nuc} } $. In contrast, when employing the $ \mathrm{NN}{-}\mathrm{N}^4\mathrm{LO}(500) {+} 3 \mathrm{N}_{ \mathrm{lnl} }^{ * } $ interaction there is a reshuffling of the amplitude strength; the residue is about half as large and the $ J = 1 $ moment of the non-residue contribution changes overall sign. Regardless, the predictions for $ \delta_{\text{NS}} $ are largely unaffected and differ absolutely by only $ \sim 2 \times 10^{ -4 } $. In Fig.~\ref{sec:results:fig:gW-box-b_momentum_distribution}, we show the decomposition of the $ \square_{ \gamma W }^{b} ( E_e ) $ integrand, for the $ i T^{ \mathrm{mag} }_{ J = 1 } \otimes i T^{ 5, \mathrm{el} }_{ J = 1 } $ amplitude, into the terms described in Eq.~\eqref{sec:theory:eq:gW-box_expansion}. The terms are plotted at fixed electron energy $ E_e \approx m_e $ over the virtual gauge boson $3$-momentum and come from the calculation with the $\text{NN-N}^{4}\text{LO}(500) + 3\text{N}_{\text{lnl}}^{*}$ interaction. Pictured is the combination of the electron residue and Wick contributions calculated up to $ \mathcal{O} ( E_e ) $, shown as dash-dotted, and the Compton residue contribution, shown as dashed, respectively labelled by \textit{expansion} and \textit{residue}. Despite the obvious discontinuities in the individual contributions at $ \magnitude{q} = \magnitude{q}_{ \mathrm{max} } $, the sum is continuous over the $3$-momentum, as expected.

%%%%
% deltaNS spread
%%%%
\begin{figure}[!t]
\captionsetup{justification=raggedright}
\centering\includegraphics[width=\linewidth]{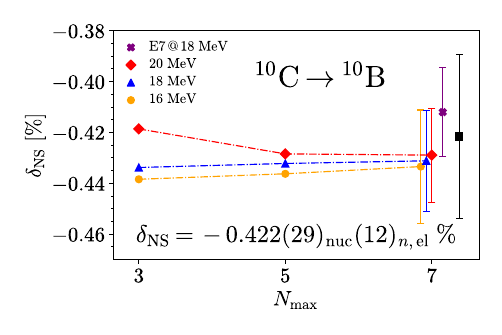}
\caption{\label{sec:results:fig:deltaNS_convergence} Sequences of $\delta_{\text{NS}}$ evaluations in the NCSM with $ N_{ \mathrm{max} } = 3,\, 5,\, 7 $ model space truncations and frequencies $\hbar \Omega = 16 - 20 \, \text{MeV} $. The $E_{7}$ point is obtained with the $\text{NN-N}^{4}\text{LO}(500) + 3\text{N}_{\text{lnl}}^{*}$ interaction; all other points are generated with the $\text{NN-N}^{4}\text{LO}(500) + 3\text{N}_{\text{lnl}}$ interaction. Error bars are as described in the text.}%
\end{figure}

In Fig.~\ref{sec:results:fig:deltaNS_convergence}, we present NCSM evaluations of $ \delta_{ \mathrm{NS} } $ over a range of increasing $ N_{ \mathrm{max} } = 3 - 7 $ configuration space sizes. Variation of the oscillator frequency characterizing the many-body basis ($ \hbar \Omega $) has been performed with the $\text{NN-N}^{4}\text{LO}(500) + 3\text{N}_{\text{lnl}}$ interaction while a single frequency was chosen for use with the $\text{NN-N}^{4}\text{LO}(500) + 3\text{N}_{\text{lnl}}^{*}$ interaction. For brevity, we quote the uncertainties entering the $ \delta_{ \mathrm{NS} } $ extractions and leave the finer details to the Supplemental Material~\cite{SupMat} (see also Ref.~\cite{Epelbaum:2014efa} therein). For a sequence of $ N_{ \mathrm{max} } $ calculations at a fixed frequency, the error bars on the final point are determined from the (i) many-body basis truncation (ii) multipole expansion truncation (iii) single-nucleon dipole form factors~\cite{Phys.Rev.C.70.068202, Seng:2018qru} and (iv) subtraction of the elastic part of the free neutron $ \gamma W $-box~\cite{Cirigliano:2022yyo, Gorchtein:2023srs}. The prediction for $ \delta_{ \mathrm{NS} } $ in Fig.~\ref{sec:results:fig:deltaNS_convergence} (black square) is obtained from averaging the two interaction calculations at $ \hbar \Omega = 18 \, \mathrm{MeV} $. The nuclear error is obtained from the quadrature combination of the partial model errors (PMEs), given by (i) $-$ (iii), along with the errors from (v) chiral expansion truncation (vi) choice of the oscillator frequency and (vii) nuclear shadowing effects, labelled by $ \chi $, $ \Omega $ and ``sh'', respectively.

Thus, we predict the nuclear-structure-dependent RC to the $^{10}$C $\to$ $^{10}$B superallowed decay rate to be
\begin{equation}\label{sec:results:eq:deltaNS_prediction}
    \delta_{ \mathrm{NS} } = - 0.422 \, (14)_{ \mathrm{PME} } \, (4)_{ \Omega } \, (9)_{ \chi } \, (24)_{ \mathrm{sh} } \, (12)_{ n, \mathrm{el} } \; \% \ \ .
\end{equation}
By construction, the uncertainty from subtraction of $ \big( \square_{ \gamma W }^{ b, n } \big)_{ \mathrm{el} } $ is totally anti-correlated to the one in the single-nucleon RC, $ \Delta_{ R }^{ V } = 0.02479 \, (12)_{ n, \mathrm{el} } \, (14)_{ n, \mathrm{inel} } \, (10)_{ \mathrm{hi} } $. This value is taken from Ref.~\cite{Gorchtein:2023srs}, with errors from the elastic (el) and inelastic (inel) parts of the single-nucleon box diagram, as well as higher-order QCD+QED corrections (hi). It is hence natural to quote the full RC
\begin{equation}\label{sec:results:eq:hadronic_RC_prediction}
    \Delta_{ R }^{ V } + \delta_{ \mathrm{NS} } = 0.02057 \, (29)_{ \mathrm{nuc} } \, (14)_{ n, \mathrm{inel} } \, (10)_{ \mathrm{hi} } \ ,
\end{equation}
as the $ ( n, \mathrm{el} ) $ uncertainty drops out in the sum. Our prediction for $ \delta_{ \mathrm{NS} } $ agrees with the value $ \delta_{ \mathrm{NS} } = -0.400 (50) \; \% $ quoted in Ref.~\cite{Hardy:2020qwl} with an overall $ 1.6 \mathrm{x} $ reduction in the total uncertainty. This is achieved despite accounting for the nuclear shadowing uncertainty not previously considered in calculations of this kind. It further disagrees with $ \delta_{ \mathrm{NS} } = -0.347 (35) \% $ quoted in Ref.~\cite{Hardy:2014qxa}, confirming the importance of the nuclear corrections due to quasi-elastic nucleon excitations pointed out in Refs.~\cite{Seng:2018qru, Gorchtein:2018fxl}.

All in all, this work provides a strategy for modern, systematically improvable calculations of general electroweak RCs in nuclei and paves the way for a more precise determination of $ V_{ud} $. Future pathways to further improving the precision of these calculations include a deeper understanding of the EFT connection of SM processes to nucleon-level processes~\cite{Cirigliano:2024msg,Cirigliano:2024rfk}, the treatment of higher-body currents at the many-body level and a more sophisticated treatment of the high-energy contributions to the nuclear structure functions. In particular, follow-up studies on nuclear shadowing effects would greatly benefit the extraction of quantities like $ \delta_{ \mathrm{NS} } $ which rely on electroweak physics that probes hadronic structure over vast energy scales. Additionally, as the extraction of $ V_{ud} $ is not limited to theory alone, future plans to improve theoretical precision serve as motivation for a new experimental determination of the $ {}^{10}\mathrm{C} $ branching ratio, which has historically suffered from large uncertainty. In fact, already in contemplation is a new measurement program using superconducting tunnel junctions to detect $^{10}$B recoils in analogy with the successful $ {}^{7}\mathrm{Be} $ electron capture measurements~\cite{PhysRevLett.125.032701, Leach:2021bvh,Carney:2022pku}. Moreover, the methodology explored in this work is readily applicable to the superallowed decays of $ {}^{14}\mathrm{O} $, $ {}^{18}\mathrm{Ne} $ and $ {}^{22}\mathrm{Mg} $, which will be studied in follow-up works, and can be easily generalized to other electroweak processes such as parity-violating electron-nucleus scattering.

%%%%
% acknowledgments and references
%%%%

\begin{acknowledgments}

We thank Vincenzo Cirigliano, Wouter Dekens, Martin Hoferichter, Emanuele Mereghetti, Oleksandr Tomalak, Doron Gazit and Ryan Plestid for many inspiring discussions. This work was supported by the NSERC Grant No. SAPIN-2022-00019. TRIUMF receives federal funding via a contribution agreement with the National Research Council of Canada. Computing support came from an INCITE Award on the Summit and Frontier supercomputers of the Oak Ridge Leadership Computing Facility (OLCF) at ORNL and from the Digital Research Alliance of Canada. M. Gorchtein acknowledges support by EU Horizon 2020 research and innovation programme, STRONG-2020 project under grant agreement No 824093, and by the Deutsche Forschungsgemeinschaft (DFG) under the grant agreement GO 2604/3-1. The work of C.-Y.S. is supported in part by the U.S. Department of Energy (DOE), Office of Science, Office of Nuclear Physics, under the FRIB Theory Alliance award DE-SC0013617, and by the DOE grant DE-FG02-97ER41014. We acknowledge support from the DOE Topical Collaboration ``Nuclear Theory for New Physics'', award No. DE-SC0023663.

\end{acknowledgments}

%%%%
% bibliography
%%%%

%

%%%%
% Supplemental material
%%%%

\setcounter{page}{1}
\renewcommand{\thepage}{Supplemental Information -- S\arabic{page}}
\setcounter{table}{0}
\renewcommand{\thetable}{S\,\Roman{table}}
\setcounter{equation}{0}
\renewcommand{\theequation}{S\,\arabic{equation}}
\setcounter{figure}{0}
\renewcommand{\thefigure}{S\,\arabic{figure}}

%%%%
% supplemental material
%%%%

\newpage
\section{Supplemental Material}

% figure of the gW-box diagram

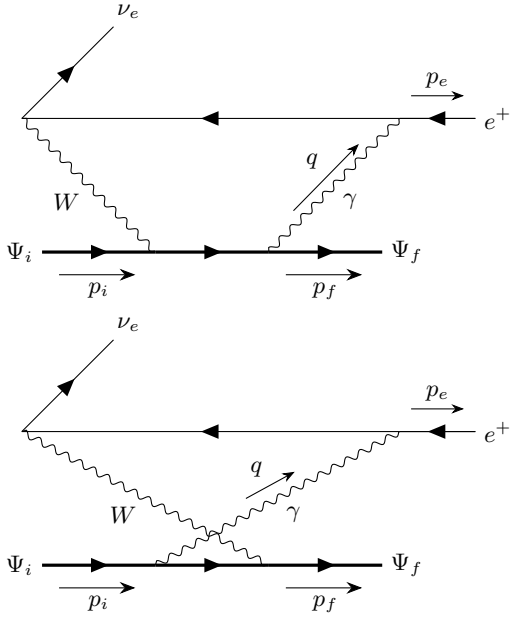
\begin{figure}[t]
\captionsetup{justification=raggedright}
\centering
\begin{tikzpicture}[arrowlabel/.style={/tikzfeynman/momentum/.cd, arrow shorten=#1, arrow distance=1.5mm}]
\begin{feynman}
    % vertices involving neutrino and electron
    \vertex (n0) { $ \nu_{ e } $ };
    \vertex [ below left=2cm of n0 ] (e1);
    \vertex [ right=5cm of e1 ] (e2);
    \vertex [ right=1cm of e2 ] (e3) { $ e^{ + } $ };
    % vertices involving nuclei
    \vertex [ below right=2.5cm of e1 ] (N2);
    \vertex [ left=of N2 ] (N1) { $ \Psi_{ i } $ };
    \vertex [ right=of N2 ] (N3);
    \vertex [ right=of N3 ] (N4) { $ \Psi_{ f } $ };
    
    \diagram* {
        % neutrino and electron lines
        (e3) -- [ fermion ] (e2) -- [ fermion ] (e1) -- [ fermion ] (n0),
        (e2) -- [ momentum=$ p_{ e } $ ] (e3),
        % nuclear lines
        (N1) -- [ fermion, very thick, momentum'=$ p_{ i } $ ] (N2) -- [ fermion, very thick ] (N3) -- [ fermion, very thick, momentum'=$ p_{ f } $ ] (N4),
        % boson lines
        (N2) -- [ boson, edge label=$ W $ ] (e1),
        (e2) -- [ boson, edge label=$ \gamma $, momentum={ [arrowlabel=0.75] $ q $ } ] (N3);
    };
\end{feynman}
\end{tikzpicture}
\begin{tikzpicture}[
arrowlabel/.style={/tikzfeynman/momentum/.cd, arrow shorten=#1,arrow distance=2.0mm}, arrowlabel/.default=0.4
]
\begin{feynman}
    % vertices involving neutrino and electron
    \vertex (n0) { $ \nu_{ e } $ };
    \vertex [ below left=2cm of n0 ] (e1);
    \vertex [ right=5cm of e1 ] (e2);
    \vertex [ right=1cm of e2 ] (e3) { $ e^{ + } $ };
    % vertices involving nuclei
    \vertex [ below right=2.5cm of e1 ] (N2);
    \vertex [ left=of N2 ] (N1) { $ \Psi_{ i } $ };
    \vertex [ right=of N2 ] (N3);
    \vertex [ right=of N3 ] (N4) { $ \Psi_{ f } $ };
    
    \diagram* {
        % neutrino and electron lines
        (e3) -- [ fermion ] (e2) -- [ fermion ] (e1) -- [ fermion ] (n0),
        (e2) -- [ momentum=$ p_{ e } $ ] (e3),
        % nuclear lines
        (N1) -- [ fermion, very thick, momentum'=$ p_{ i } $ ] (N2) -- [ fermion, very thick ] (N3) -- [ fermion, very thick, momentum'=$ p_{ f } $ ] (N4),
        % boson lines
        (N3) -- [ boson, edge label=$ W $ ] (e1),
        (e2) -- [ boson, edge label=$ \gamma $, momentum={ [arrowlabel=0.6] $ q $ } ] (N2);
    };
\end{feynman}
\end{tikzpicture}
\caption{\label{sec:SuMa:fig:gammaWbox} The nuclear $ \gamma W $-box diagrams.}%
\end{figure}

This Supplemental material collects information not displayed in the core manuscript that falls into one of the following categories:
\begin{itemize}
    \item It is available in the quoted references and is reproduced for the reader's convenience
    \item It involves technical details of the \textit{ab initio} calculation which do not alter the main discussions
    \item It is needed for a complete analysis of the $ \delta_{ \mathrm{NS} } $ uncertainties, but is independent of the \textit{ab initio} calculation
\end{itemize}

%%%%
% subsection on the [ gW-box ](E_e) terms from Wick rotation
%%%%
\subsection{Terms in the nuclear box diagram}

Here we provide the analytic expressions for the three terms on the right-hand-side of Eq.~\eqref{sec:theory:eq:gW-box_split} which stem from the nuclear $ \gamma W $-box diagram in Fig.~\ref{sec:SuMa:fig:gammaWbox}; approximate forms have been previously derived in Refs.~\cite{Seng:2022cnq, Gorchtein:2023naa} but generalized here to retain the full $ E_e $- and $ m_e $-dependence. First, the Wick-rotated term reads
\begin{align}\label{sec:SuMa:eq:boxWick}
    & \big( \square_{ \gamma W }^b \big)_{ \mathrm{Wick} } ( E_e ) = - \frac{ \alpha }{ \pi M } \int_{ 0 }^{ \infty } d\magnitude{q} \ \magnitude{q}^2 \int_{ - \infty }^{ \infty } \frac{ d\nu_{ E } }{ 2\pi } \frac{ M_W^2 }{ M_W^2 + Q^2 } \nonumber \\
    & \qquad \times \frac{ 1 }{ Q^2 } \frac{ T_3 ( i\nu_E, \magnitude{q} ) }{ f_+ \nu_E } \Bigg\{ \frac{ 2 ( \magnitudeexp{ p_{e} }{ 2 } \magnitude{q}^2 + E_e^2 \nu_E^2 ) - i Q^2 \nu_E E_e }{ 4 \magnitudeexp{ p_{e} }{ 3 } \magnitude{q} } \nonumber \\
    & \qquad \times \ln \Bigg( \frac{ - 2 i E_e \nu_E - Q^2 + 2 \magnitude{ p_{e} } \magnitude{q} }{ - 2 i E_e \nu_E - Q^2 - 2 \magnitude{ p_{e} } \magnitude{q} } \Bigg) - \frac{ i \nu_E E_e }{ \magnitudeexp{ p_{e} }{ 2 } } \Bigg\} \ , \nonumber \\
\end{align}
where $ \magnitude{ p_{e} } = | \vec{p}_e | $ and $ Q^2 = \nu_E^2 + \magnitude{q}^2 $. Next, the residue contribution from the electron propagator reads
\begin{align}\label{sec:SuMa:eq:boxe}
    & \big( \square_{ \gamma W }^{ b } \big)_{ \mathrm{res}, e } ( E_e ) = \frac{ i \alpha }{ 2 \pi M } \int_{ 0 }^{ 2 \magnitude{ p_{e} } } d\magnitude{q} \ \magnitude{q}^2 \int_{ \magnitude{q} / 2 \magnitude{ p_{e} } }^{ 1 } dx \ \frac{ 1 }{ \nu_e^2 - \magnitude{q}^2 } \nonumber \\
    & \qquad \qquad \times \frac{ \frac{ \nu_e \magnitude{q} E_e }{ \magnitude{ p_{e} } } x - \magnitude{q}^2 }{ \sqrt{ E_e^2 - 2 \magnitude{ p_{e} } \magnitude{q} x + \magnitude{q}^2 } } \ \frac{ T_3 ( \nu_e, \magnitude{q} ) }{ f_+ \nu_e } \quad ,
\end{align}
where
\begin{align}
    \nu_e = E_e - \sqrt{ E_{ e }^2 - 2 \magnitude{ p_{e} } \magnitude{q} x + \magnitude{q}^2 } \ .
\end{align}
Adding Eqs.~\eqref{sec:SuMa:eq:boxWick},~\eqref{sec:SuMa:eq:boxe} and expanding the result in powers of $ E_e $ (in the $ m_e = 0 $ limit) yields
\begin{align}\label{sec:SuMa:eq:regularexpand}
    & \big( \square_{ \gamma W } \big)_{ \mathrm{Wick} } ( E_e ) + \big( \square_{ \gamma W } \big)_{ \mathrm{res}, e } ( E_e ) \nonumber \\
    & \qquad \quad = \boxminus_{ 0 } + \boxminus_{ 1 } E_e + \mathcal{O}( E_{ e }^2 ) \quad ,
\end{align}
where the coefficients $\boxminus_{0,1}$ may be found in Ref.\cite{Gorchtein:2023naa}. We note that, while convenient, such expansion is not necessary. Finally, the residue contribution from the Compton amplitude $ T_3 $ due to a given state $ k $ in the low-lying nuclear spectrum of $ {}^{10}\mathrm{B} $ reads
\begin{align}\label{sec:SuMa:eq:boxT3}
    & \big( \square_{ \gamma W }^{ b } \big)_{ \mathrm{res}, T_3 } = \frac{ i \alpha }{ \pi M } \int_{ 0 }^{ \magnitude{q}_{ \mathrm{max} } } d\magnitude{q} \ \magnitude{q}^2 \ \frac{ \mathrm{Res} \; T_3( \nu_k, \magnitude{q} ) }{ f_+ } \nonumber \\
    & \qquad \quad \times \frac{ 1 }{ \magnitude{q}^2 - \nu_k^2 } \ \Bigg\{ \frac{ 2 \magnitudeexp{ p_{e} }{ 2 } \magnitude{q}^2 + \nu_k E_e ( \nu_k^2 - 2 E_e \nu_k - \magnitude{q}^2 ) }{ 4 \nu_k \magnitudeexp{ p_{e} }{ 3 } \magnitude{q} } \qquad \nonumber \\
    & \qquad \quad \times \ln \Bigg\vert \frac{ \nu_{ k }^2 - 2 E_{ e } \nu_{ k } - \magnitude{q}^2 + 2 \magnitude{ p_{e} } \magnitude{q} }{ \nu_{ k }^2 - 2 E_e \nu_k - \magnitude{q}^2 - 2 \magnitude{ p_{e} } \magnitude{q} } \Bigg\vert - \frac{ E_e }{ \magnitudeexp{ p_{e} }{ 2 } } \Bigg\} \ , \nonumber \\
\end{align}
where
\begin{equation}
    \mathrm{Res} \; T_3( \nu_k, \magnitude{q} ) = \lim_{ \nu \rightarrow \nu_k } ( \nu - \nu_k ) T_3( \nu, \magnitude{q} ) \ \ ,
\end{equation}
with
\begin{equation}
    \nu_k ( \magnitude{q} ) = M_k - M_f + \frac{ \magnitude{q}^2 }{ 2 M_k } \ \ ,
\end{equation}
and
\begin{equation}
    \magnitude{q}_{ \mathrm{max} } = \sqrt{ M_f^2 - M_k^2 } \ \ .
\end{equation}
%

%%%%
% subsection on the multipole expansion formalism for SM currents
%%%%
\subsection{Multipole expansion and dependence on nucleon form factors}\label{sec:multipole_expansion}
%%%%
% subsection on the multipole expansion formalism for SM currents
%%%%

Connection of the SM currents appearing in the Compton amplitude in Eq.~(\ref{sec:theory:eq:gW_box}) to the effective nucleon-level one-body operators compatible with nuclear many-body theory is accomplished via the tried and true multipole expansion formalism. Comprehensive reviews on the topic may be found in Refs.~\cite{World.Sci.2004, Comp.Phys.Comm.179.2008, Atom.Nuc.Tables.23.1979}. For the reader's convenience, we quote the definition of the relevant nucleonic operators appearing in Eq.~(\ref{sec:theory:eq:T3}).
\begin{align}\label{sec:SuMa:eq:Tel}
    & T^{ \mathrm{el} }_{ J M ; \kappa } ( q, \magnitude{x} ) \\
    & \quad = \frac{ \magnitude{q} }{ m_N } \bigg\{ F_{ 1 } ( q^2 ) \, \Delta'_{ J M } ( \magnitude{q} \magnitude{x} ) + \frac{ 1 }{ 2 } \mu (q^2) \, \Sigma_{ J M } ( \magnitude{q} \magnitude{x} ) \bigg\} \, \mathcal{I}_{ \kappa } \nonumber
\end{align}
\begin{align}\label{sec:SuMa:eq:Tmag}
    & i T^{ \mathrm{mag} }_{ J M ; \kappa } ( q, \magnitude{x} ) \\
    & \quad = \frac{ \magnitude{q} }{ m_N } \bigg\{ F_{ 1 } ( q^2 ) \, \Delta_{ J M } ( \magnitude{q} \magnitude{x} ) - \frac{ 1 }{ 2 } \mu (q^2) \, \Sigma'_{ J M } ( \magnitude{q} \magnitude{x} ) \bigg\} \, \mathcal{I}_{ \kappa } \nonumber
\end{align}
\begin{align}\label{sec:SuMa:eq:T5el}
    & i T^{ 5, \mathrm{el} }_{ J M ; \kappa } ( q, \magnitude{x} ) = - F_{ A } (q^2) \, \Sigma'_{ J M } ( \magnitude{q} \magnitude{x} ) \; \mathcal{I}_{ \kappa }
\end{align}
\begin{align}\label{sec:SuMa:eq:T5mag}
    & T^{ 5, \mathrm{mag} }_{ J M ; \kappa } ( q, \magnitude{x} ) = F_{ A } (q^2) \, \Sigma_{ J M } ( \magnitude{q} \magnitude{x} ) \; \mathcal{I}_{ \kappa }
\end{align}
In these expressions $ \kappa \in \{ 0, 3, \pm \} $ and the non-spherical isospin operator (depending on the isospin coordinates of the underlying SM current at hand) is given by
\begin{equation}
    \mathcal{I}_{ 0 } = \mathbb{I} \qquad \quad \mathcal{I}_{ 3 } = \tau_{ 3 } \qquad \quad \mathcal{I}_{ \pm } = \tau_{ \pm } \quad .
\end{equation}
We are consistent with the form factor independent electroweak operator basis defined in Ref.~\cite{Atom.Nuc.Tables.23.1979} which assumes the vanishing of second-class currents. The combination of the form factor independent pieces into the full multipole operators is accomplished by applying the dipole form of the single-nucleon form factors; discussion pertaining to the error attributed to this choice is discussed in the following sections.

%%%%
% subsection with derivation of shadowing effects
%%%%
\subsection{The nuclear shadowing effect}

In this section, we outline the details from which we obtain Eq.~\eqref{sec:theory:eq:deltaNSmod} and the subsequent estimation of the nuclear shadowing uncertainty. As pointed out in this Letter, the full nuclear box diagram $ \square_{ \gamma W }^{ b, \mathrm{nuc} } $ appearing in Eq.~\eqref{sec:theory:eq:box_subtraction} includes contributions from all nucleonic excitations and non-nucleonic, multi-particle hadronic (MPH) excitations. However, the piece that we compute using the \textit{ab initio} NCSM and leading order currents, denoted $ \big( \square_{ \gamma W }^{ b, \mathrm{nuc} } \big)_{ \mathrm{ab} } $, only captures the former. The remaining MPH intermediate states unaccounted for in the nuclear model represent contributions to the inelastic part of the $ \gamma W $-box. It is then prudent to decompose the $ \gamma W $-box into elastic and inelastic contributions as
\begin{equation}\label{sec:shadowing:eq:gWbox_separation}
    \square_{ \gamma W }^{ b,n } = \big( \square_{ \gamma W }^{ b,n } \big)_{ \mathrm{el} } + \big( \square_{ \gamma W }^{ b,n } \big)_{ \mathrm{inel} } \quad .
\end{equation}
The good agreement in the deep inelastic region between the experimentally-determined Gross-Llewellyn Smith (GLS) sum rule~\cite{Gross:1969jf} on iron target~\cite{Kataev:1994rj,Kim:1998kia} and the theory prediction up to $ \mathcal{O} ( \alpha_{ s }^{ 4 } ) $~\cite{Larin:1991tj, Baikov:2010iw, Baikov:2010je} indicates that the pQCD contribution to $\big( \square_{ \gamma W }^{ b,n } \big)_{ \mathrm{inel} } $ arising from high loop momenta, i.e., $ Q^2 > 2 ~ \mathrm{GeV}^2 $, is largely unaffected by nuclear structure and hence does not enter $ \delta_{ \mathrm{NS} } $.

What remains is to account for the MPH contribution to the single-nucleon box diagram in the regime of $ Q^2 < 2 \ \mathrm{GeV}^2 $. Averaging over several dispersion relation analyses~\cite{Gorchtein:2023srs} yields
\begingroup
\begin{align}\label{sec:shadowing:eq:MPHvalue}
    \big( \square_{ \gamma W }^{ b,n } \big)_{ \mathrm{inel} }^\mathrm{MPH} &\equiv \big( \square_{ \gamma W }^{ b,n } \big)_{ \mathrm{inel} } \big( Q^2 < 2 \ \mathrm{GeV}^2 \big) \\
    &\approx 0.59(6) \times 10^{-3} \nonumber \quad .
\end{align}
\endgroup
Note that we do not include the result from the recent lattice calculation of $ \square_{ \gamma W }^{ b,n } $~\cite{Ma:2023kfr} since the separation shown in Eq.~\eqref{sec:shadowing:eq:gWbox_separation} was not performed. We will now use this estimate to reconstruct the missing piece of the $ \square_{ \gamma W }^{ b, \mathrm{nuc} } $.

In literature, it has always been assumed that $ \big( \square_{ \gamma W }^{ b,n } \big)_{ \mathrm{inel} } $ remains unmodified in the nuclear medium. Generally speaking, this assumption is unwarranted and so we consider rather
\begin{equation}
    \big( \square_{ \gamma W }^{ b,n } \big)_{ \mathrm{inel} } \ \rightarrow \ \big( \square_{ \gamma W }^{ b,n } \big)_{ \mathrm{inel} } + \delta \big( \square_{ \gamma W }^{ b,n } \big)_{ \mathrm{sh} } \quad ,
\end{equation}
where $ \delta \big( \square_{ \gamma W }^{ b,n } \big)_{ \mathrm{sh} } $ denotes the shadowing-driven modification of the MPH contribution. This translates into the following decomposition of $ \delta_{ \mathrm{NS} } $ displayed in Eq.~\eqref{sec:theory:eq:deltaNSmod} of the main text, which we repeat here for convenience:
\begin{align}
    \delta_{ \mathrm{NS} } &= 2 \bigg\{ \big( \square_{ \gamma W }^{ b, \mathrm{nuc} } \big)_{ \mathrm{ab} } - \big( \square_{ \gamma W }^{ b, n } \big)_{ \mathrm{el} } + \delta \big( \square_{ \gamma W }^{ b,n } \big)_{ \mathrm{sh} } \bigg\} \quad . \nonumber
\end{align}
We emphasize that the effect of nuclear shadowing is not calculable using \textit{ab initio} approaches and so, instead, we provide a conservative estimate of its size which is then folded into the theoretical uncertainty of $ \delta_{ \mathrm{NS} } $.

To infer the approximate size of nuclear shadowing effects we utilize available data on the parity-even structure function $ F_{ 2 }^{ \mathrm{nuc} } $. From $ {}^{9} \mathrm{Be} $ to $ {}^{12} \mathrm{C}$, the upper bound on the nuclear modification of MPH contributions varies between $ 10 \% $ and $ 20 \% $ (see, e.g., Figs. 22 and 51 of Ref.~\cite{Kopeliovich:2012kw} and references therein). Thus, we take $ 20 \% $ of the central value of $ \big( \square_{ \gamma W }^{ b,n } \big)_{ \mathrm{inel} }^\mathrm{MPH} $ in Eq.~\eqref{sec:shadowing:eq:MPHvalue} as an upper bound on the uncertainty in $ \square_{ \gamma W }^{ b, \mathrm{nuc} } $ incurred from nuclear shadowing effects. That is,
\begin{align}
    \big\vert \delta \big( \square_{ \gamma W }^{ b,n } \big)_{ \mathrm{sh} } \big\vert
    &\lesssim 1.2 \times 10^{-4} \quad .
\end{align}

Some implicit assumptions in the above treatment, e.g., the similarity between the shadowing effects in P-even and P-odd structure functions, should be further scrutinized with an in-depth theoretical study using a more suitable framework such as the Glauber-Gribov approach~\cite{glauber1959, Gribov:1968jf}. This will be pursued in a future work.

%%%%
% subsection with more details on the error analysis
%%%%
\subsection{Detailed error analysis}

In this section we provide the finer details regarding the various nuclear modelling errors entering into the evaluations given in Fig.~\ref{sec:results:fig:deltaNS_convergence} as well as Eqs.~(\ref{sec:results:eq:deltaNS_prediction}) and~(\ref{sec:results:eq:hadronic_RC_prediction}), sans the nuclear shadowing effects which have been thoroughly explored in the previous section. A precise summary of the error breakdown is provided in Table.~\ref{tab:deltaNS_uncertainties}.

First and foremost, due to the computational complexity of nuclear many-body calculations we encounter three types of truncation in these calculations: (i) the truncation of the many-body HO basis in the NCSM, controlled by the parameter $ N_{ \mathrm{max} } $ (ii) the truncation of the multipole expansion, controlled by the parameter $ J_{ \mathrm{max} } $ and (iii) the truncation of the chiral expansion. Note that truncation of the chiral expansion affects both the modelling of the nuclear Hamiltonian and the SM current operators whose respective chiral orders are not taken into account consistently. Consequently, their associated truncation errors are estimated independently. Furthermore, the nuclear Hamiltonian is SRG evolved while the current operators are not. This approximate treatment of the fully consistent SRG evolution leads to an additional source of uncertainty from neglecting higher-order corrections to the current; their impact must be estimated on top of the chiral truncation. Overall, this amounts to estimating a total uncertainty coming from our modelling of the nuclear Hamiltonian and SM currents. We will now characterize each of these uncertainties one by one.

Beginning with the $ N_{ \mathrm{max} } $ truncation which controls the many-body basis size, it is clear that in the limit $ N_{ \mathrm{max} } \rightarrow \infty $ we approach the exact NCSM result. At finite $ N_{ \mathrm{max} } $, we estimate the truncation error $\epsilon_{ N_{ \mathrm{max} }}$ by taking the absolute difference of consecutive $ N_{ \mathrm{max} } $ evaluations of $ \delta_{ \mathrm{NS} } $. In addition, we use the residual dependence on the HO frequency $\Omega$ to estimate the error incurred, $\epsilon_{\Omega}$, by choosing the frequency $\hbar\Omega=18 \ \mathrm{ MeV}$ when evaluating $\delta_{ \mathrm{NS} }$. It is estimated by taking the maximum absolute difference between varied frequency calculations in the vicinity of $ \hbar \Omega = 18 \ \mathrm{MeV} $.

On the other hand, as illustrated in Figs.~(3) and~(4) of Ref.~\cite{Phys.Rev.C.100.064315}, the integrated amplitudes of odd and even electroweak multipoles present an approximately exponential decay in the rank $ J $ of the considered operator structure. Moreover, such an observation is reinforced by the well-understood asymptotic behaviour of the Bessel functions which decay factorially in $ J $. By fitting a two-parameter exponential to the integrated amplitudes for a given operator structure in $ T_3 $, we thus estimate the associated error $ \epsilon_{ J_{ \mathrm{max} } } $ by summing over the extrapolated values for the contributions with $ J > J_{ \mathrm{max} } = 3 $, i.e., the amount of missing strength due to truncation of the multipole expansion. As the multipoles are expected to decay at least exponentially in $ J $, this yields a reasonable estimate of the missing strength.

Lastly, error is introduced due to truncation of the chiral expansion in the Hamiltonian $ \epsilon_{ \chi } $ and in the modelled electroweak currents $ \epsilon_{ \mathrm{hc} } $. A well motivated approach to the estimation of chiral interaction uncertainties $ \epsilon_{ \chi } $ at the many-body level is discussed in Ref.~\cite{Epelbaum:2014efa} and applied in Ref.~\cite{Phys.Lett.B.845.138156} with the NCSM. It involves fully consistent calculations at each order in the chiral expansion which, purely as a result of the cost of these calculations, we are not able to perform. Instead, we may still reasonably estimate the effects of the truncation by varying the chiral interaction in use. To this end, we have considered the so-called $E_7$ interaction which includes an additional sub-leading contact interaction in the three-nucleon sector. The final result is taken as the average of those two calculations (at a consistent oscillator frequency) and its dispersion is used as the additional uncertainty $\epsilon_{\chi}$.

\begin{table}[t!]
\captionsetup{justification=raggedright}
  \begin{center}
    \begin{tabular}{lc|cr} 
      \hline
      \hline
      $^{10}$C $\to$ $^{10}$B & \hspace{0.25cm} This work \hspace{0.25cm} & \hspace{0.25cm} Ref.~\cite{Hardy:2020qwl} \hspace{0.25cm} & $\times 10^{-4}$\\
      \hline
      % \hline
      $ \epsilon_{ J_{ \mathrm{max} } } $ & $ 0.1 $ & $ 3.3 $ & $\epsilon_{\delta_{\mathrm{NS},A}}$ \\
      $ \epsilon_{ M_{ \mathrm{scale} } } $ & $ 0.04 $ & $ 3.5 $ & $\epsilon_{\delta_{\mathrm{NS},B}}$ \\
      $ \epsilon_{ N_{ \mathrm{max} } } $ & $ 0.1 $ & $ 1.5 $ & $\epsilon_{\delta_{\mathrm{NS},E}}$ \\
      $ \epsilon_{ \mathrm{hc} } $ & $ 1.4 $ & $/$ & \\
      \hline
      % \hline
      $ \epsilon_{ \mathrm{PME} } $ & $ 1.4 $ & $ / $ & \\ % $\epsilon_{\mathrm{PME}}$ \\
      $ \epsilon_{ \Omega} $ & $0.4$ & $/$ & \\ % $\epsilon_{\Omega}$ \\
      $ \epsilon_{ \chi} $ & $0.9$ & $/$ & \\ % $\epsilon_{H_{\mathrm{nuc}}}$ \\
      $ \epsilon_{ \mathrm{sh} } $ & $2.4$ & $/$ & \\ % $\epsilon_{\mathrm{sh}}$ \\
      \hline
      % \hline
      $\epsilon_{\mathrm{nuc}}$ & $2.9$ & $/$ & \\ % $\epsilon_{\mathrm{nuc}}$ \\
      % \hline
      $\epsilon_{n,\mathrm{el}}$ & $1.2$ & $/$ & \\ % $\epsilon_{n,\mathrm{el}}$ \\
      % \hline
      % $\epsilon_{\mathrm{had}}$ & & $0$ & $\epsilon_{\mathrm{had}}$ \\
      \hline
      % \hline
      $\epsilon_{\delta_{\mathrm{NS}}}$ & $3.1$ & $5.0$ & $\epsilon_{\delta_{\mathrm{NS}}}$ \\
      \hline
      \hline
    \end{tabular}
    \caption{\label{tab:deltaNS_uncertainties} List of different uncertainties accounted for in the $\delta_{\mathrm{NS}}$ calculation discussed in this Letter. For comparison, we also provide the uncertainties considered in Ref.~\cite{Hardy:2020qwl}. Different sub-groups correspond to different degree of aggregation of the uncertainties.}%
  \end{center}
\end{table}

What remains is to estimate the error $ \epsilon_{ \mathrm{hc} } $ from lacking higher-order currents and the approximate SRG evolution, which are most conveniently estimated together. To do so, we rely entirely on previous literature studies which have systematically explored such effects on the vector electromagnetic current (see Fig.~3 of Ref.~\cite{Phys.Rev.Lett.126.102501} for $ \mathrm{M1} $ transitions) and the axial part of the axial-vector weak current (see Table~IV and Fig.~8 in the Supplemental Material of Ref.~\cite{Gysbers2019} for $ \mathrm{GT} $ transitions, as well as Fig.~2 of Ref.~\cite{Phys.Rev.C.109.065501} for $ \mathrm{GT} $-like transitions) in light nuclei, albeit primarily at zero momentum transfer. While individually significant effects, an important observation made in all of the cited references is the partial cancellation between the effect of consistent SRG evolution of the operators and the inclusion of higher-order currents. In the end, the combined effect is reduced to the level of a few percent. Detailed in Ref.~\cite{Gysbers2019}, the total shift was shown to be approximately equivalent to a phenomenological re-scaling of the current coupling constants which, in our case, is equivalent in phenomenology to a re-scaling of the single-nucleon form factors, e.g., $ F_1 $, $ \mu $ and $ F_A $. Further noted in such studies is that the effects of higher-order contributions to the vector electromagnetic and axial weak currents are not uncorrelated, i.e., the effects observed in the same system (where the literature studies exist) are similar in magnitude but systematically opposite in sign. Thus, both enhancement of the vector electromagnetic current and quenching of the axial weak current must be considered simultaneously in the amplitudes; we perform a parametric re-scaling of the form factors and extract $ \epsilon_{ \mathrm{hc} } $ from the variation in $ \delta_{ \mathrm{NS} } $.

The expected enhancement and quenching of our operator structures is estimated \textit{ad hoc} by considering the percentage change between amplitudes with no SRG evolution and no higher-order current contributions to those with both such treatments, as reported in the aforementioned works. For the exact numbers, see Fig.~3 and Table~I of Ref.~\cite{Phys.Rev.Lett.126.102501} for $ \mathrm{M1} $ transitions, as well as Table~IV and Fig.~8 in the Supplemental Material of Ref.~\cite{Gysbers2019} for $ \mathrm{GT} $ transitions. We find that the enhancement in the vector electromagnetic current in $ \mathrm{M1} $ transitions is approximately $ 2 - 4 \% $ and the quenching in the axial weak current for $ \mathrm{GT} $ transitions is approximately $ 3 - 5 \% $. As available studies are limited to such operators, these numbers are applied universally regardless of the operator structure at hand. Consequently, we vary the single-nucleon form factors of the vector electromagnetic current according to
\begin{equation}\label{sec:errors:eq:vector_rescale}
    F_{ 1 } \rightarrow \Tilde{ F }_{ 1 } = c_{ V } \, F_{ 1 } \qquad \quad \mu \rightarrow \Tilde{ \mu } = c_{ V } \, \mu \nonumber \quad ,
\end{equation}
with $ c_{ V } \in \big\{ 1.02, \, 1.04 \big\} $, and similarly the single-nucleon form factors of the axial weak current according to
\begin{equation}\label{sec:errors:eq:axial_rescale}
    F_{ A } \rightarrow \Tilde{ F }_{ A } = c_{ A } \, F_{ A } \nonumber \quad ,
\end{equation}
with $ c_{ A } \in \big\{ 0.97, \, 0.95 \big\} $. We then extract the uncertainty explicitly as
\begin{equation}
    \epsilon_{ \mathrm{hc} } = \sup_{ c_V , \, c_A } \big\vert \bar{\delta}_{ \mathrm{NS} }( c_V, \, c_A ) - \delta_{ \mathrm{NS} } \big\vert \quad ,
\end{equation}
where $ \delta_{ \mathrm{NS} } $ is the prediction of a given calculation and $ \bar{\delta}_{ \mathrm{NS} } ( c_V , \, c_A ) $ is a function on the $ c_V $-$ c_A $ parameter space.

The above discussion characterizes all truncation uncertainties entering the \textit{ab initio} many-body calculation of $ \delta_{ \mathrm{NS} } $. The final item to be discussed is the treatment of the $4$-momentum dependence of the electroweak multipoles which, as identified in the mutlipole expansion section, are decomposed into an electroweak operator basis under the assumption of vanishing second-class currents~\cite{Atom.Nuc.Tables.23.1979}. The remaining model dependence consists of the $4$-momentum character of the single-nucleon form factors utilized for electromagnetic and weak processes.

For a general form factor appearing in Eqs.~(\ref{sec:SuMa:eq:Tel}) -- (\ref{sec:SuMa:eq:T5mag}), we use the standard dipole form factor approximation
\begin{equation}
    F ( q^2 ) = F ( 0 ) \bigg[ 1 - \frac{ q^2 }{ M_{ \mathrm{scale} }^2 } \bigg]^{ -2 } \quad ,
\end{equation}
\\where in this context $ M_{ \mathrm{scale} } $ refers to either the vector dipole mass $ M_V $ or the axial dipole mass $ M_A $. This approximation is uncontrolled in the sense that it is merely recognized as a convenient fitting procedure for a given form factor's momentum distribution. Nevertheless, we can conservatively estimate the error $ \epsilon_{ M_{ \mathrm{scale} } } $ arising from such an approximation by varying the corresponding dipole mass to span the entire range of more sophisticated form factor fitted to reproduce experimental data. Proceeding in this way, the vector dipole mass is varied in a range of $ 800 \text{ -- } 1000 \ \mathrm{MeV} $ to cover the range of predictions from the high-quality Pad{\'e} fits performed in Ref.~\cite{Phys.Rev.C.70.068202} whereas the axial dipole mass is varied in a range of $ 1.09 \text{ -- } 1.270 \ \mathrm{GeV} $ as recommended in Ref.~\cite{Seng:2018qru}. The combination of $ \epsilon_{ N_{ \mathrm{max} } } $, $ \epsilon_{ J_{ \mathrm{max} } } $, $ \epsilon_{ \mathrm{hc} } $ and $ \epsilon_{ M_{ \mathrm{scale} } } $ make up the partial model error $ \epsilon_{ \mathrm{PME} } $ given in Table~\ref{tab:deltaNS_uncertainties}.
%{\color{magenta} Kelly's parametrization is 20 years old and is obsolete. Why are we using it? Besides, in a one-parameter form like the dipole the scale is uniquely related to the radius which is certainly known better than to 20\%. See \url{https://arxiv.org/pdf/2312.08694} for some recent polemics. The values for $r_M$ lie within 0.77-0.85 fm or 0.81(4)fm. In any case, the respective uncertainty is negligible.}.
At the moment, the errors discussed in this section represent the largest contributions to the uncertainty in the $ \delta_{ \mathrm{NS} } $ prediction. Future systematic improvements to the formalism are envisioned, for example, improvements to the theoretical calculation of $ \delta_{ \mathrm{NS} } $ could be achieved by inclusion of higher-body currents and relativistic corrections or simply by using a more precise extractions of form factors. Finally, we note that most of the developments discussed here can be extended to other electroweak radiative corrections.

\end{document}